\documentclass[reprint,twocolumn,showpacs,nofootinbib,showkeys,longbibliography,
amssymb,preprintnumbers,superscriptaddress,notitlepage,pra]{revtex4-1}

\usepackage{graphicx}
\usepackage{latexsym}
\usepackage{amsfonts}
\usepackage[usenames]{color}
\usepackage{url,hyperref}
\usepackage{bm}
\usepackage{natbib}
\hypersetup{colorlinks=true, citecolor=blue, linkcolor=red, urlcolor=blue}
\definecolor{dred}{rgb}{0.75,0.08,0}

\newcommand{\vp}{\varphi}
\newcommand{\bb}{\mathbf}
\newcommand{\nn}{\nonumber\\}
\newcommand{\beq}{\begin{equation}}
\newcommand{\eeq}{\end{equation}}
\newcommand{\bed}{\begin{displaymath}}
\newcommand{\eed}{\end{displaymath}}
\def\bea{\begin{eqnarray}}
\def\eea{\end{eqnarray}}
\newcommand{\veps}{\varepsilon}
\newcommand{\unit}{\hat\mathbf}
\newcommand{\bnab}{\bm\nabla}

\begin{document}

\title{Vacuum-excited surface plasmon polaritons}
\author{Wade Naylor}
\email[Email: ]{naylor@phys.sci.osaka-u.ac.jp}
\affiliation{International College and Department of Physics, Interdisciplinary Research Building, Osaka University, Toyonaka, Osaka 560-0043, Japan}

\begin{abstract}
We separate Maxwell's equations for background media that allow for both electric and magnetic time-dependence in a generalized Lorenz gauge. In a process analogous to the dynamical Casimir effect (DCE) we discuss how surface plasmon polaritons (SPP)s can be created out of vacuum, via the time-dependent variation of a dielectric and magnetic insulator at a metal interface for TM and TE branches, respectively. We suggest how to extend currently proposed DCE experiments to set up and detect these excitations. Numerical simulations (without any approximation) indicate that vacuum excited SPPs can be of a similar magnitude to the photon creation rate in such experiments. Potential benefits of detecting vacuum excited SPPs, as opposed to DCE photons, are that parametric enhancement does not require a sealed cavity in the axial direction and the detection apparatus might be able to use simple phase matching techniques. For the case of constant permeability, $\mu$, TM branch SPPs and photons do not suffer from detuning and attenuation like TE photons.
\end{abstract}

\begin{figure}[t]
\vspace{-1.5cm}
\hspace{-7.0cm}
\end{figure}

\preprint{OU-HET-781/2013}
\date{\today}

\maketitle


\section{Introduction}

\par Particle creation via the Schwinger-effect \cite{Schwinger:1951nm}, in expanding universes \cite{Parker:1968mv} or from black hole evaporation \cite{Hawking:1974rv} all have yet to be confirmed.\footnote{This excludes analog set ups, e.g., see \cite{Nation:2012}. In particular for graphene there are some promising proposals \cite{Allor2008,Mostepanenko2013} to observe a $2+1$ dimensional Schwinger effect.} However a related effect known as the dynamical Casimir effect (DCE), first discussed by Moore \cite{moore:2679}, is in experimental reach. For the parametric oscillations of a mirror contained in a cavity the number of photons created is proportional to $\sinh^2 (2\omega t \,v/c)$, e.g., see \cite{Dodonov:2010zza}, where $v$ is the wall velocity and $c$ is the speed of light. To overcome the fact that the mechanical properties of the material usually imply $v/c\ll 1$, there have been proposals other than mechanical oscillations. Modulating a dielectric medium using a laser also leads to particle creation by varying the optical path length of the cavity, e.g., see \cite{UhlmanPRL2004}. There are experiments in progress in three-dimensional centimeter-sized (microwave) cavities \cite{Agnesi:2009jpc}, where a laser is used to modulate the surface conductivity. 
Other methods use illuminated superconducting boundaries \cite{Segev:2007} and recently time varied inductance effects in one-dimensional quantum circuits have already demonstrated vacuum squeezing \cite{Wilson:2012, Paraoanu_PNAS110}. Rotating analogs have also been investigated \cite{MagrebPRL108}.

\par
\begin{figure}[t]
\centering
\scalebox{0.35}{\includegraphics{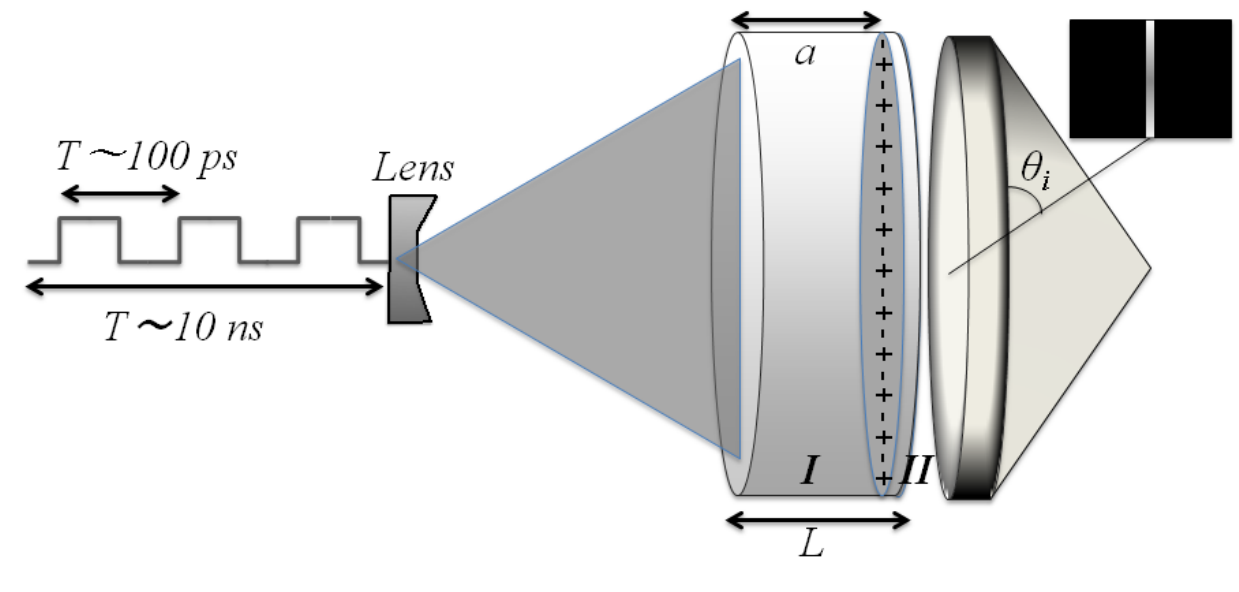}}
\vspace{-0.3cm}
\caption{A pulse laser train of order $10 - 100$ pulses  (repeating $\sim 10$ ms) uniformly irradiates (via a lens) a semiconductor/metal (SM) interface of radius $R$, composed of a dielectric of thickness $a$, region $I$ at a thin metal (e.g., silver) interface of thickness $(L-a)\ll L$, region $II$. Vacuum excited SPPs ({\tiny$-\,+\,-\,+\,-\,+$}) are detected using a phase matched prism placed to the right of region $II$.  
}
\label{exp}
\end{figure}

 In this article we explore the possibility of the creation of vacuum excited surface plasmon polaritons (SPPs) and how they might be detected. In Fig.~\ref{exp} a pulsed laser of an appropriate frequency can be used to vary the time dependence of a dielectric. The crystal can also be placed in a superconducting cavity (not shown), to suppress thermal excitations and lend to parametric enhancement of the photon creation rate. SPPs are by definition damped modes of oscillation in the perpendicular direction and therefore only affected by the transverse dimensions. This means that only the transverse dimensions need to be enclosed to obtain parametric enhancement, which might be a potential benefit experimentally. 

\par A telltale signature of the creation of SPPs would be {\it an increase in emitted power at position $\theta_i$} when coupled to a phase-matched prism at one end of the semiconductor-metal (SM) interface.\footnote{Here we consider a semiconductor semispace. Usually in plasmonics an insulator-metal (IM) interface is assumed, e.g., see \cite{Maier}.} This is in stark contrast to the usual SPP generation method/detection, e.g., see \cite{Maier}, where a decrease in emitted power at $\theta_i$ occurs via illumination of the prism. As well as SPPs, TE and TM photon pair creation is also expected; however these are at different frequencies and would not couple to the prism, and would further require a sealed cavity (we discuss more on detection methods later).

\par The fact that SPPs can be excited from vacuum fluctuations besides photons is much like SPPs in the static Casimir force \cite{Intravaia:2005zz} for a metal-insulator-metal (MIM) heterostructure. We essentially generalize this idea to the dynamical case at first for the simpler single interface: a semiconductor-metal (SM) interface and show that the time modulation of a dielectric leads not only to two-photon pair creation  processes, but to vacuum excited SPPs that are comparable to the photon creation rate, under certain conditions. This may well have important consequences for experiments currently trying to detect pair created photons. Theoretically, a dynamical Casimir effect for single a interface arises from the analogy that a single moving boundary emits DCE radiation, even though there is no Casimir force \cite{Dodonov:2010zza}.

\par The outline of the article is as follows. In the next section (Sec. \ref{method}) we give details on the theory behind time-dependent surface plasmons, while in Sec. \ref{genplasma} we use generalized plasma model to obtain analytic expressions for the SPP dispersion relations. In Sec. \ref{part} we discuss how to numerically evaluate the particle creation rate via a full separation of variables without any approximation, while in Sec. \ref{detect} we propose possible detection schemes. and we conclude in Sec. \ref{conc}. Extra material is left for Appendices: relating to Maxwell's equations for time dependent dielectrics, App. \ref{timedepMax}; the Hertz vectors approach to separation, App. \ref{Hertz}; and comparing the exact separation of variables with the instantaneous basis approach, App. \ref{instbas}.

\section{Theory} 
\label{method}

\par Our theoretical starting point is the following Lagrangian (from which Maxwell's equations can be derived):
\beq
{\cal L} = \frac 1 2   \veps(t)  \left({\partial\over \partial t}\Phi\right)^2 -\frac 1 2{1\over\mu(t)} (\nabla \Phi )^2 - \frac 1 2 m^2(t)\Phi^2
\label{Lag}
\eeq
($\veps_0=\mu_0=1$). In the above we assume that the electric permittivity and magnetic permeability are time dependent, but piecewise constant in space: $\nabla \mu=\nabla \veps=0$. $\Phi$ represents a TM field with generalized Neumann BCs in a cavity and the TE case (swapping $\veps \leftrightarrow \mu$) is represented by $\Psi$ with Dirichlet  BCs, e.g., see \cite{Naylor:2012ia}. This Lagrangian is useful because the standard canonical Hamiltonian can be constructed \cite{Yamamoto:2011pra}, where the mass term, $m^2(t)$, represents the coupling of light to a time-dependent boundary ($m^2$ can also arise from considering an electron plasma).

\par A convenient way to separate Maxwell's equations is using Hertz vectors; developed by Nisbet \cite{Nisbet:1957} for non-dispersive inhomogeneous media. However, here we generalize to the case of a constant isotropic, but time-dependent medium. It is  possible to show, see App. \ref{timedepMax}, that Maxwell's equations separate as
\bea
\veps(t) \partial_t ({\mu(t) \partial_t  {\mathbf \Pi}_e}) -\nabla^2 { {\mathbf \Pi}_e} &=& 0~,
\nn
\mu(t) \partial_t ({\veps(t)\partial_t  {\mathbf \Pi}_m}) -\nabla^2 { {\mathbf \Pi}_m} &=& 0~,
\label{Maxwell}
\eea
where we use a generalized Lorenz gauge (also discussed in App. \ref{timedepMax}): 
\beq
\mu(t) \partial_t (\veps(t) A_0) + \bm\nabla\cdot \mathbf A = 0,
\eeq
cf. \cite{Nisbet:1957,Bei:2011} and see Eq. (\ref{lorenz}). In the above we have assumed both a zero permanent polarization and magnetization ($\bb P_0=\bb M_0=0$) as well as zero bulk charges and currents ($\rho=0, \,\bb J =0$), although these can also be included in the Hertz method. Note the Lagrangian in Eq.~(\ref{Lag}) leads to the equations of motion for  $\Pi_e$ in Eq.~(\ref{Maxwell}) ($\Pi_m$ is obtained by swapping $\mu \leftrightarrow \veps$ in Eq.~(\ref{Lag})), see App. \ref{timedepMax}. This approach generalizes other work \cite{UhlmanPRL2004,Bei:2011} which considered only time-dependent $\veps$ or $\mu$. Further work for non-dispersive, inhomogeneous, conducting and time-dependent media: $\veps(\bb r,t)$ and $\mu(\bb r,t)$,  will be presented elsewhere.

\par Before  quantizing the SPP modes we first need to find the classical solutions for a single interface (between two media) that lead to SPPs. 
Writing the electric and magnetic fields in terms of Hertz vectors:
\bea
{\mathbf E}
&=&{1\over \veps}\bnab\times(\bnab\times {\mathbf \Pi_e})-\mu_0\bnab\times {\partial_t\mathbf \Pi_m}~,
\nn
{\mathbf B} &=& \mu \bnab\times {\partial\mathbf \Pi_e\over\partial t}+
\mu_0\bnab\times(\bnab\times {\mathbf \Pi_m})~,
\label{fields}
\eea
allowing one to easily isolate TE and TM modes, see App. \ref{Hertz}. 


\par In what follows we take two half spaces in the $\unit z$-direction, where region 1 (the semiconductor slab) is a semiconductor (I) and region 2 a metal (M) like silver, creating an SM interface.
In our proposed set up $\veps_1(t)$ varies from a minimum to maximum value and $\veps_2<0$ remains constant (although for now it will be left more general). To make explicit the utility of the Hertz vector method we shall consider the radial propagation of SPPs in a cylindrical cavity with coordinates $(\rho,\theta,z)$, sectional radius $\rho=R$ and length $L$, see Fig.~\ref{exp}. 

\par Using the Hertz potentials  (\ref{Maxwell}) and assuming from symmetry that $\bb \Pi_{ m}= \Psi~\hat \mathbf z$ for TE, and $\bb \Pi_{ e}=\Phi~\hat \mathbf z$ for TM modes, the separation of variables:
\bea
\Psi(\bb x,t)=\sum_{\bb l} \psi_{\bb l}(\bb x) q^m_{\bb l}(t)~,
\nn
\Phi(\bb x,t)=\sum_{\bb l} \phi_{\bb l}(\bb x) q^e_{\bb l}(t)~,
\label{sep}
\eea
with $\bb l = (n,p,l)$, leads to the following wave equation, e.g., see \cite{PedrosaPRL2009}:
\beq
\bb\nabla^2 \psi_{\bb l}(\bb x)+\veps(t)\mu(t) \omega_{\bb l}^2(t)\psi_{\bb l}(\bb x)=0~.
\label{nabla}
\eeq
with the replacement $\psi_{\bb l} \to \phi_{\bb l}$ for TM modes. 
They satisfies the standard orthonormality conditions:
\beq
\int_{-\infty,0}^{\infty, L} d^3{\bb x}\,\psi_{\bb l}(\bb x)\psi_{\bb n}(\bb x)=(\psi_{\bb l},\psi_{\bb n})=\delta_{\bb l\bb n}
\label{ortho}
\eeq 
where the $(-\infty,\infty)$ bounds on the integral are for SPPs and that with $(0,L)$ are for the photon branch (see later). We then find that the time dependent part satisfies
\bea
\ddot {q}^m_{\bb l} + {\dot\veps \over \veps} \dot {q}^m_{\bb l} + \omega_{m \bb l}^2 q^m_{\bb l}=0~,
\label{TEtimeQ}
\\
\ddot {q}^e_{\bb l} + {\dot\mu \over \mu} \dot {q}^e_{\bb l} + \omega_{e \bb l}^2 q^e_{\bb l}=0~.
\label{TMtimeQ}
\eea
where the superscripts $m, e$ are for TE and TM modes respectively. Importantly, we see that for setups with only dielectrics present $\mu_{I,II}=$ constant then the TM mode functions are simple Mathieu like equations with natural frequency given by $\omega_{e \bb l}$. On the other hand for TE modes, the presence of $\dot \veps$ and $\dot q_m$ leads to a detuning via $\tilde\omega_{m \bb l}$, see Eq. (\ref{mathieu}), and also losses for $\dot\veps>0$ \cite{PedrosaPRL2009}.

\par The conjugate momentum ${\cal P}_m=\partial {\cal L}/\partial t$ can be found from Eq.~(\ref{Lag}) along with the separation ansatz and orthonormality relations implying
\beq
{\cal P}_m(\bb x, t) = \veps(t)\sum_{\bb l}  \psi_{\bb l}(\bb x) p^m_{\bb l}(t)
\eeq 
and via a Legendre transform we find the time dependent Hamiltonian for each mode $\bb l$ (TE):
\beq
H^m_{\bb l} = \veps^{-1}
{(p^m_{\bb l})^2\over 2}+ {\veps\over  2} \omega_{\bb l}^2(t)(q^m_{\bb l})^2
\label{Ham}
\eeq
where the conjugate mode momentum is defined by\footnote{The $x,y$ dependence of the mode functions decouples and can be written in terms of the index $\bb l \to l$ from now onwards.} 
\beq
p^m_{\bb l} = \dot q^m_{\bb l}.
\eeq 

\par Given the ETCRs: $[\hat q_{\bb l}, \hat p_{\bb n}]=i\delta_{\bb l\bb n}$, we get back the equation of motion, Eq.~(\ref{TEtimeQ}), from the above Hamiltonian.  
A similar analysis applies to TM modes: $\Phi,{\cal P}_e$ and hence we can quantize each degree of freedom $(\Phi, \Psi)$. In the above we rescaled the coordinates as $q^m_{\bb l} \to \veps^{-1/2} q^m_{\bb l}$ for TE and would need $q^e_{\bb l} \to \mu^{-1/2} q^e_{\bb l}$  for TM modes (see later). In terms of these creation and annihilation operators we see squeezing terms in the Hamiltonian \cite{PedrosaPRL2009}.

\subsection*{SPP and Photon Branches}

\par To investigate SPPs for a single interface the ansatz:
\beq
\label{plas}
\Phi_{\rm sp}(\bb x, t) = 
\left\{
\begin{array}{ccc}
A_{1} e^{\kappa_{1l} (z-a)}r_{\rm np}(\bb x_\bot) \,,  & ~  z<a\,;~\veps_1,~\mu_1\,, \\
 A_{2} e^{-\kappa_{2l}(z-a)}r_{\rm np}(\bb x_\bot)\,, &~ z>a\,;~\veps_{2},~\mu_{2}\,,
\end{array}
\right. 
\eeq
leads to the following `time-dependent' dispersion relations: 
\beq
\bb k_\bot^2-\veps_1 \mu_1 {\omega_{\bb l}^2\over c^2}= \kappa_{1l}^2\,,\qquad \bb k_\bot^2-\veps_{2} \mu_{2} {\omega_{\bb l}^2\over c^2}= \kappa_{2l}^2
\label{sppdisp}
\eeq
where Eq.~(\ref{Maxwell}) was used in each region. In cylindrical coordinates the transverse Laplacian is
defined by
\beq
-\bnab_{\bot}^2 r_{\bb k_\bot} = \bb k_\bot^2 r_{\bb k_\bot}
\eeq 
with eigenvalue $\bb k_\bot^2$. 
In DCE experiments the slab is usually bounded by a cavity
 (not depicted in Fig.~\ref{exp}) where 
\beq
r_{\rm np}(\bb x_\bot)={1\over \sqrt\pi}{1\over R J_{n+1}(x_{\rm np})}J_n\Big(x_{\rm np}\frac \rho R \Big)e^{in\theta}~,
\label{cylr}
\eeq
$x_{\rm np}$ is the $p$th root of $J_{n}(x)=0$ \cite{Jackson} and the fundamental cavity mode is $x_{01}=2.4048$. 
For a cavity bounding the SM interface in the transverse directions we would have $R_{sp} \leq R=2.5$ cm and hence 
$ (\bb k^{\rm sp}_\bot)^2=(x_{np}/R_{\rm sp})^2\sim {\cal O} (1)$, this depends on the value of $q$, cf. Eq. (\ref{qval}). For the photon branch we always have $(\bb k^{\rm ph}_\bot)^2=(x_{np}/R)^2$ for a bounded cavity. Note in either case the mode functions are orthornormal: $(r_{\ell n},r_{np})=\delta_{\ell p}$.

\par Standard boundary conditions at an interface:
\beq
( {\mathbf D}_{2}-{\mathbf D}_1)\cdot {\unit z}=0~, 
\qquad
{\unit z}\times( {\mathbf E}_{2}-{\mathbf E}_1) =0
\eeq
\cite{Jackson} then imply $A_1=A_{2}$ and 
\beq
{\kappa_1(t)\over \veps_1(t)} + {\kappa_{2}(t)\over\veps_{2}(t)} = 0
\label{junct}
\eeq
which requires that each dielectric be of opposite sign to generate SPPs \cite{Maier}. 
 Eliminating the $z$-dependent $\kappa_i$ we then find the following `electric' dispersion relation:
\beq
\label{spTE}
k_\bot = |\bb k_\bot| = {\omega_\bot^{\rm sp}\over c} \sqrt{\veps_1\veps_{2}\over\veps_1 +\veps_{2}}\times\left(\veps_1 \mu_{2}-\veps_{2}\mu_1\over\veps_1 -\veps_{2}\right)
\eeq
With $\mu_1=\mu_{2}$ we get the standard result
\beq
\label{spTEsimple}
(\omega_\bot^{\rm sp})^2 = {k_\bot^2 c^2} \left({1\over \veps_1}+{1\over \veps_2}\right)~,
\eeq 
where here we allow for time-dependent dielectrics, possibly in either region $I$ and $II$ and $\bot = (n,p)$ because for SPPs the axial direction $l$ is redundant. In Sec. \ref{genplasma} we 
will use a plasma-type model to obtain more detailed analytic properties of the above dispersion relation. 


\par It is also worth mentioning that magnetic SPPs exist for TE modes \cite{Nesterenko:2012fk}. Using the TE components of the Hertz vectors and using an equation like Eq.~(\ref{plas}) for $\Psi(\bb x, t)$ along with
\beq
( {\mathbf B}_{2}-{\mathbf B}_1)\cdot {\unit z}=0 ~,
\qquad
{\unit z}\times( {\mathbf H}_{2}-{\mathbf H}_1) =0
\eeq
lead again to $A_1=A_{2}$ but now with
\beq
{\kappa_1(t)\over\mu_1(t)} + {\kappa_{2}(t)\over\mu_{2}(t)} = 0.
\eeq
As also discussed in \cite{Nesterenko:2012fk}, SPPs can exist for TE modes as long as for example, $\mu_1 <0,\,\mu_2>0$, which can be achieved using split ring resonators, e.g., see \cite{Maier}: using fabricated metamaterials. Finally using $\bb \Pi_m$ in Eq.~(\ref{Maxwell}) leads to the `magnetic' dispersion relation:
\beq
\label{spTM}
k_\bot = {\omega_\bot^{\rm sp}\over c} \sqrt{\mu_1\mu_{2}\over\mu_1 +\mu_{2}}\times\left(\mu_1 \veps_{2}-\mu_{2}\veps_1\over\mu_1 -\mu_{2}\right)~.
\eeq
This result can be obtained from the `electric' sector by swapping $\veps_i \leftrightarrow \mu_i$, and simplifies when $\veps_1=\veps_2$ to
\beq
(\omega_\bot^{\rm sp})^2 = {k_\bot^2 c^2} \Big({1\over \mu_1}+{1\over\mu_2}\Big)~.
\eeq
As also discussed in \cite{Nesterenko:2012fk} this implies that TE modes can sustain surface plasmons; however, the material needs to be a fabricated metamaterial.

\par  To compare vacuum excited SPPs with some experimental proposals for photon creation using semiconductor slabs, e.g., see \cite{Dodonov:2010zza}, we will  also consider TM modes in a slab of width $(L-a)$, placed in a cylindrical cavity of length $L$ (not shown). These have the following orthonormal mode functions for the TM photon Hertz scalar:
\beq
\Phi_{\rm ph}(\bb r, t) = 
\left\{
\begin{array}{ccc}
A_1{\cos\,(k_{1l} z)} r_{\rm np}(\bb x_\bot),   &  0<z<a, \\
 A_2\cos\,(k_{2l}(L-z))r_{\rm np}(\bb x_\bot), & a<z<L.
\end{array}
\right. 
\eeq
where (using the same TM interface conditions as before) we find following transcendental 
equation
\beq
 \frac{ k_{1l}  \tan(k_{1l}a) }{\veps_1(t)}
 = \frac{ k_{2l}  \tan(k_{2l}[a-L]) }{\veps_2(t)}
 \label{trans}
\eeq
 for the eigenvalues. This agrees with the result  in \cite{UhlmanPRL2004} but can be derived with the minimum of effort using Hertz vectors and generalized to arbitrary transverse section. Note that the photon dispersion relation (in this case for a cylindrical section) at any given time in regions $i=1,2$:
\beq
{\omega_{i\bb l}^{\rm ph}(t) } ={c\over \veps_i(t)}\sqrt{k_{il}^2(t) + \left(\frac{x_{\rm np}}{R}\right)^2}
\label{wTM}
\eeq 
must be equal at the interface implying equivalence of the dispersion relations:
\beq
\frac 1 {\veps_1} \left(k_{1l}^2+ \left(\frac{x_{\rm np}}{R}\right)^2\right)= \frac 1 {\veps_2}\left(k_{2l}^2+ \left(\frac{x_{\rm np}}{R}\right)^2\right).
\label{inter}
\eeq 
Note this dispersion relation is the complex conjugate of that in Eq. (\ref{sppdisp}): $k=i\kappa$.
Both this constraint and the eigenvalue relation, Eq.~(\ref{trans}), must be simultaneously satisfied \cite{UhlmanPRL2004}. For slab thicknesses with $L-a\ll L$ (or for $a\ll L$) one can further show \cite{UhlmanPRL2004} that even for quite large variations in the dielectric constant the approximate solution to Eq. (\ref{trans}) is (for $l>0$)
\beq
k_{1l}(t)= \Big({l\pi \over L}\Big)\Big( 1 - {a\over L}\Big[{\veps_1(t)\over\veps_2} -1\Big]
\left(\frac{x_{\rm np}}{R}\right)^2 \left(\frac{L}{l \pi}\right)^2
\Big)~
\label{ktmapprox}
\eeq
(note TE modes at ${\cal O} (a/L)$ are still unperturbed free modes \cite{UhlmanPRL2004}). 

\par Here we ignore the  zero modes, $l=0$, as previous work \cite{Naylor:2012ia} for plasma sheets showed they are not excited; however, see discussion in \cite{Uhlmann:2004} for dielectrics. In this case the interface constraint, Eq. (\ref{inter}), suggests that $k_{i0}=0$ ($i=1,2$) implies $\veps_1=\veps_2$ and is therefore only satisfied for time independent (static) cases. The general case, not just for $\eta=a/L\ll 1$, will be investigated more thoroughly elsewhere. 
We  therefore assume the TM${}_{011}$ is the lowest mode and investigate the number of created particles for this and the TE${}_{111}$ fundamental mode (up to ${\cal O} (a/L)$), comparing them to that for SPPs.

\hspace{3cm}
\section{Generalized Plasma Model} 
\label{genplasma}

\subsection*{Region $II$: Time Independent}

\par To simplify our analysis we will now consider a slight generalization of the plasma model \cite{Yamamoto:2011pra} of a metal-like substance in region $II$:\footnote{We could similarly include the magnetic permeabilities: $\mu_1,\mu_2$, but for simplicity we set them to unity.}
\beq
\veps_2(\omega)=\bar\veps_2\left(1- \left({\omega_p^2\over\omega^2} \right)\right)
\label{plasma}
\eeq
where $\omega_p=ne^2/(\bar\veps_2 m_*)$ is the plasma frequency and $n$ is the number of bulk electrons and $m_*$ is the effective mass. The extra multiplicative factor arises by including a mass term in the Lagrangian, Eq. (\ref{Lag}). In our envisaged experiment $\bar\veps_2$ will be a constant, but for generality we have left it time dependent, $\bar\veps_2(t)$, in the analysis below. Note the dielectric permittivity in region $II$ takes negative values for $\omega < \omega_p$. For region $I$ we assume a semiconductor material that is modulated by laser irradiation. 

\par If we then substitute Eq (\ref{plasma}) into Eq. (\ref{spTEsimple}) (using $\omega=\omega_\bot^{\rm sp}$) we obtain a generalization of the solution found in \cite{Nesterenko:2012fk}:

\bea
(\omega_\bot^{\rm sp})^2&=&\omega_p^2 \left[\frac{1}{2} + \frac{1}{2} q^2\left({1\over \veps_1}+{1\over \bar\veps_2}\right)\right. \nn
&&- \left. \sqrt{\frac{1}{4}+ \frac{1}{4}q^4\left({1\over \veps_1}+{1\over \bar\veps_2}\right)^2+\frac 1 2 q^2
\left({1\over \bar\veps_2}-{1\over \veps_1}\right)}\right]\nn
\label{plasmadis}
\eea
where  
\beq
q={k_\bot c\over \omega_p}
\label{qval}
\eeq
and we obtain the standard result for $\bar \veps_1=\veps_2=1$ \cite{Nesterenko:2012fk}. 

\par We also find the following asymptotic behavior: 
\bea
&\omega_\bot^{\rm sp}\to {k_\bot c }\left({1\over \veps_2}\right)^{1/2} + {\cal O}(q^{3}) + \dots  & \quad  q\to 0,\nn
&&\\
 &\omega_\bot^{\rm sp}\to{\omega_p\over \sqrt 2}+ {\cal O}(q^{-2}) + \dots  & \quad  q\to\infty~,
\nonumber
\eea
which explains why for small $q$, we have a system that behaves like TE/TM modes in a 1D cavity. Hence for a given form of modulation (see below) of the dielectric, the limit $q\to 0$ leads to parametric amplification of SPPs; while $q\to \infty$ gives  no SPP production (for $\omega_p =$ constant).

\subsection*{Region $I$: Time Dependence}

\par To be more specific, in this paper we will consider two kinds of modulation of the dielectric in region $I$, given that in Eq (\ref{plasma}) for region $II$. 

\par One, an inverse sinusoidal modulation:
\beq
{1\over \veps_1}=\frac{1}{ 2} \Big({1\over\veps_{\rm 1,min}} + {1\over\veps_{\rm 1,max}}\Big) +
 \frac{1} {2} \Big({1\over \veps_{\rm 1,min}} - {1\over \veps_{\rm 1,max}}\Big) \cos(2\omega_0 t) 
\label{invsin}
\eeq
\\
that has been argued to arise from the excitation of localized electrons in a semiconductor, via laser irradiation (e.g., see \cite{Uhlmann:2004}). In this regard, we should mention that for the excitation of electrons to the conduction band, instead of using a dielectric model, such as Eq. (\ref{invsin}), the conductivity modulates by assuming the plasma frequency varies with, for example, a sin-like time dependence: $\omega_p(t)=e^2 n(t)/m_*$, where $n_s(t)\propto \sin(2\omega_0 t)$, and an equation like that in Eq. (\ref{plasma}) but instead for region $I$. This is an interesting problem but differs in that $\veps_1<0$ for certain modulations and will be left for future work (also see \cite{Naylor:2012ia} for more on plasma sheet models). In this article we will assume that $\veps_{1,min},\veps_{1,max}>0$.

\par Another way to realistically modulate the permittivity, $\veps_1(t)$, but this time sinusoidally would be to use an appropriately doped semiconductor (with two well defined energy levels within the band gap) via Rabi oscillations, e.g, see \cite{Haug_Koch}: 
\beq
{\veps_1}=\frac{1}{ 2} \Big({\veps_{\rm 1,max}} + {\veps_{\rm 1,min}}\Big) +
 \frac{1} {2} \Big({\veps_{\rm 1,max}} - {\veps_{\rm 1,min}}\Big) \cos(2\omega_0 t)
\label{sin}
\eeq
\\
In the next section we shall assume that for both cases we have $\veps_{1,min}=0.2$ and $\veps_{1,max}=3.2$ which are typical values for a germanium semiconductor.


\section{Particle Creation Rates}  
\label{part}

\par To find the number of particles created we use an alternative to the Bogoliubov method using only mode functions \cite{Kofman1997}. We start with the quantum field operator expansion in the Heisenberg representation for our TM Hertz potential:
\beq
\hat{\Phi}({\bf x},t) =\sum_{ l} 
\left[ \hat a_{\bb l} \vp_{ l}(\bb x) q_{ l}(t)+ \hat a_{ l}^{\dagger}  \vp^*_{ l}(\bb x) q^*_{ l}(t)
\right]~,
\eeq
where $\hat a_{ l}, \hat a_{ l}^{ \dagger}$ are annihilation and creation operators respectively and the mode functions $ \vp_{\bb l}(\bb x),~ q_{\bb l}(t)$ were defined in Eqs. (\ref{nabla},\ref{ortho},\ref{TEtimeQ}); here we need to impose initial conditions at $t=0$: $q_l(0)={1\over \sqrt{ 2\tilde\omega_l}}$ and $\dot q_l(0)=-i\sqrt{\tilde\omega_l\over 2}$. 
To find a separable time-dependent solution we can rescale the field as $\tilde q_{\bb l} = \veps^{1/2} q_{\bb l}$ to get an equation in Mathieu form:
\beq
\ddot {\tilde q}_{ l}  + \tilde\omega_l^2(t) \tilde q_{ l}=0~.
\label{mathieu}
\eeq 
where 
\beq
\tilde\omega_l^2=\left[\omega_{l}^2+\frac 1 4 {\dot \veps^2\over \veps^2} - \frac 1 2 {\ddot \veps \over \veps} \right]~.
\label{tildome}
\eeq
It may be worth mentioning that this equation is equivalent to a scalar potential in a curved spacetime with conformal coupling $\xi=1/6$ and scale factor $\veps(t)=a(t)$ for a Robertson-Walker spacetime \cite{Birrell}.

 \par The particle number density can be obtained directly from the energy of each mode divided by the energy $\omega_{ l}$ of each particle:
\beq
n_{l} = {\tilde\omega_l\over 2} \left( 
{|\dot{\tilde q}_l|^2\over \tilde\omega_l^2}+ |\tilde q_l|^2
\right) -\frac 1 2
\label{numbden}
\eeq
where we have subtracted off the zero point energy with units $\hbar, c=1$. Eq. (\ref{mathieu}) has a well known structure of narrow or broad resonances for certain parameters. We stress that this method has separated variables without using an instantaneous basis approximation (see App. \ref{instbas}).


\par Before numerically solving for the number of created particles, we will estimate the pair creation rate analytically. If the background field (the laser) leads to shifts in frequency near to parametric resonance:  
\beq
\omega_{\bb m}^2 (t)\sim \omega_{0\bb m}^2+\Delta  \omega_{0\bb l}^2= \omega_{0\bb m}^2(1+ \kappa \cos(\Omega_{\bb l} t)),
\eeq
where the driving frequency is chosen as  $\Omega_{\bb m}=2\omega_{0\bb m}$, where ${\bb m}=(\bot,{\bb l})$, for SPPs or photons respectively, then in the late time limit:
\beq
n_{\bb m} \approx \sinh^2 \left({\omega_{0\bb m} \kappa  t /4 } \right),
\label{late}
\eeq
which can be derived by ignoring second order time derivatives in Eq. (\ref{mathieu}) \cite{DodonovPRA47}.

\par As a simple example consider $\mu_1=\mu_2$, where the time-dependent `electric' SPP dispersion relation,  Eq.~(\ref{plasmadis}), could be varied using a laser with driving frequency, $\Omega_l=2\omega_{0l}$ for $\bar \veps_2>0$  and constant, cf. Eq. \ref{plasma}, with $\veps_1$ varying inverse sinusoidally as $\veps_{\rm min}<\veps_1(t)<\veps_{\rm max}$ then
\beq
{\veps_2 \over \veps_1(t)} \sim \chi+\kappa \cos(2\omega_{0l} t),
\eeq
where $\chi>0$ is an overall time-independent frequency shift (cf. Eq (\ref{invsin}). 
Then Eq. (\ref{late}) leads to a particle rate:
\beq
n_\bot^{\rm sp}=\sinh^2\left( k_\bot^2 c^2 {\kappa \over 4\veps_2 } t \right)~.
\eeq
This equation is also valid for the more general case of $\veps_2 <0$ not just the model discussed in Sec. \ref{genplasma}.

\par We can now compare this to the $\omega_{011}$ TM mode (the lowest frequency cylindrical mode \cite{Naylor:2012ia}) where in the limit of $(L-a)\ll L$, from Eq. (\ref{ktmapprox}) and equivalence of the dispersion relations (see below Eq. (\ref{wTM})), the photon eigenvalues shift by (to order ${\cal O}(a/L)$):
\beq
\Delta \omega_{0\bb l}^2(t)= \frac {2 x_{np}^2 c^2}{R^2 \veps_2 } \frac a {L} \left[ \frac{\veps_2}{ \veps_1(t)} - 1\right].
\eeq
Parametric enhancement for the photon branch is then achieved by choosing 
\beq
{ \veps_2 \over \veps_1(t)}\sim \chi + \kappa \cos(2\omega_{0l} t)~,
\eeq 
where for photons $\chi$ and $\kappa$ may or may not be the same as those for SPPs; however they are assumed of the same magnitude. Note the resonant frequencies are not the same: $\omega_{0\bot}\neq \omega_{0\bb l}$. This leads to:
\beq
n_{\bb l}^{\rm ph}=\sinh^2\left( {x_{np}^2 c^2 \over R^2}{ a\kappa \over 2\veps_2 L } t \right).
\eeq
Thus, for a cylindrical cavity the SPP creation rate dominates the photon rate if $k_\bot^2 \gg (x_{np}^2/R^2)(2a/L)$. For example, with a cavity of radius, $R=2.5$, cm  and length, $L=10$ cm, then for $a/L \sim {\cal O}[10^{-4}]$ and $x_{01}=2.4048$ we require that $k_\bot^2 \gg 1/25$, or $k_\bot \gg 1/5$. This is easily achieved for SPPs which have their modes bounded by a transverse section.

\begin{figure}[t]
\centering
\scalebox{0.43}{\includegraphics{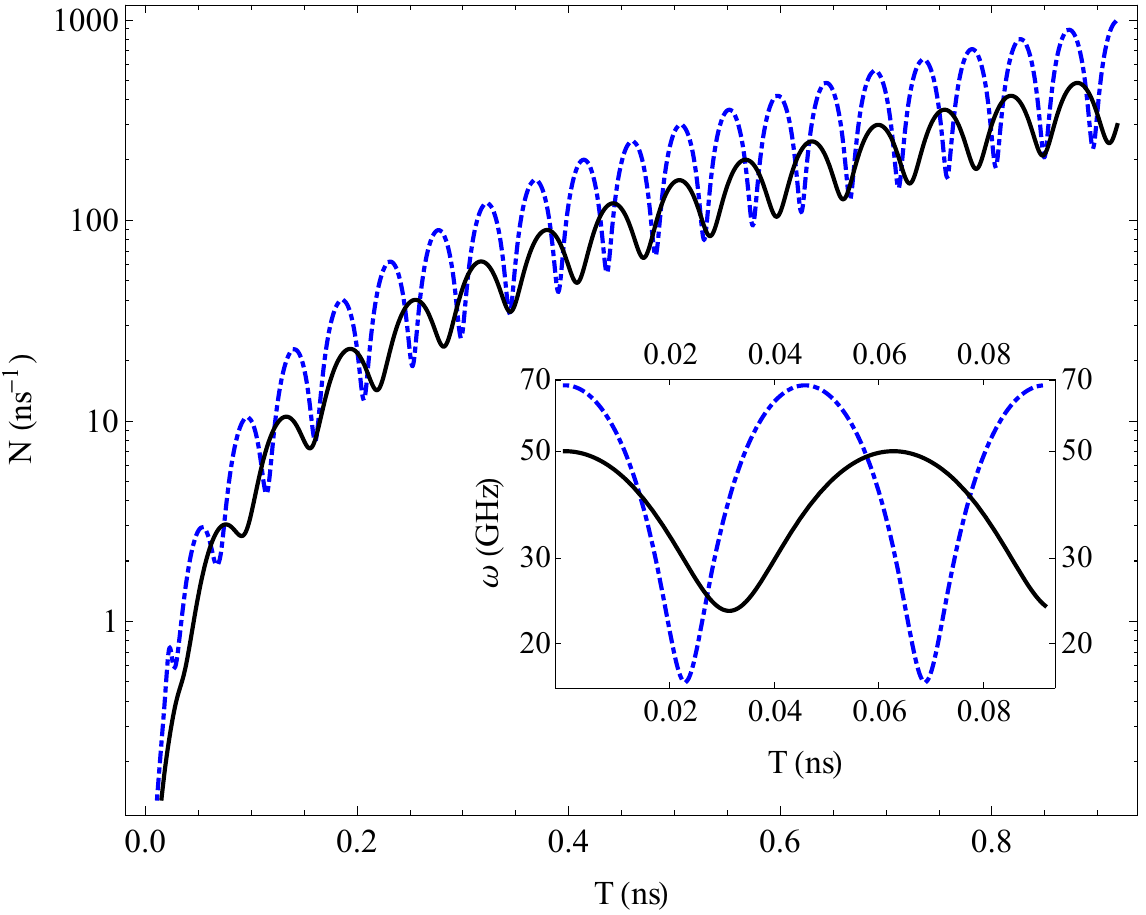}}
\\
\scalebox{0.43}{\includegraphics{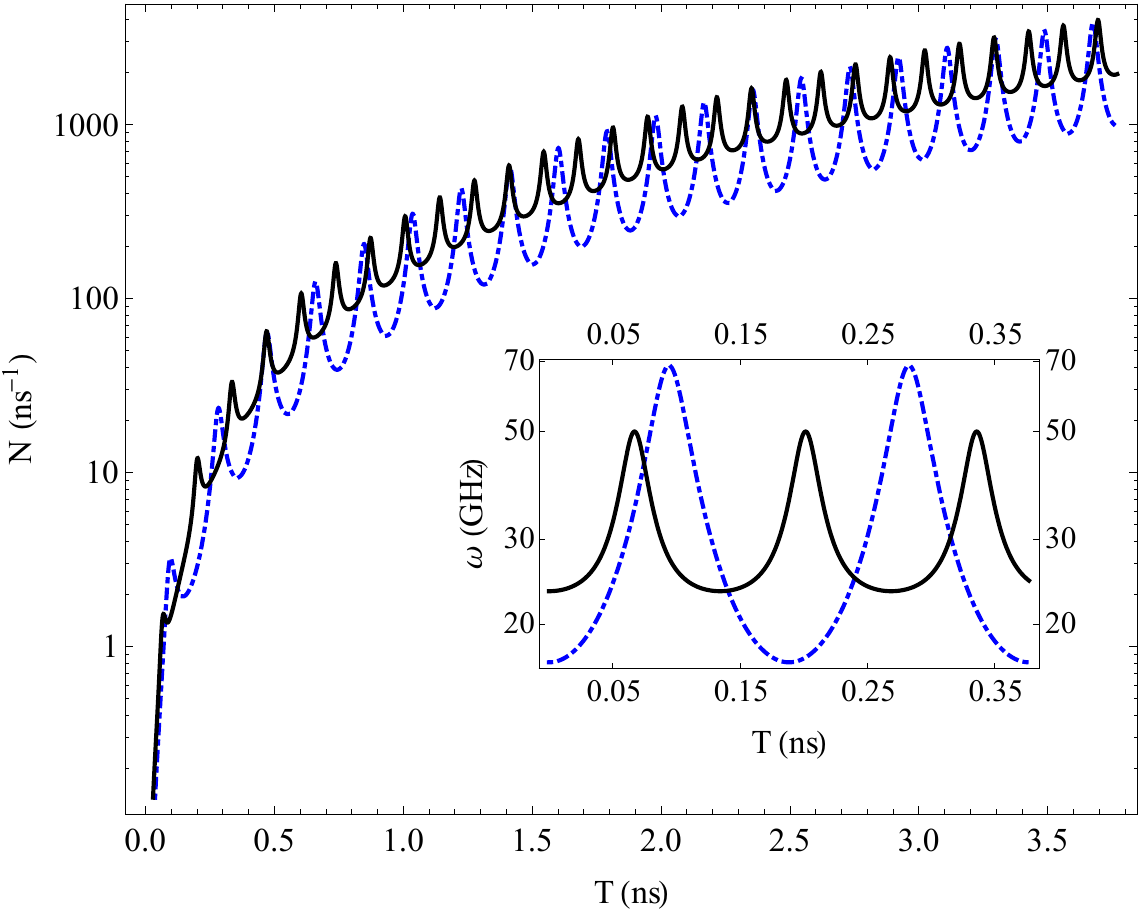}}
\caption{(Color online) Particle creation rates for SPPs (solid-dark) and 2nd fundamental TM$_{011}$ (dot-dashed-blue), for a pulse train of $~100$ pulses, $\eta=0.01$, $R_{\rm sp}=R=2.5$ cm, region $II$ with $\bar \veps_2=1.0$ and $\omega_p=1.5\times 10^{15}$ s${}^{-1}$ for silver. In region $I$ we have: a) an inverse sinusoidal variation (upper panel) and b) a sinusoidal variation (lower panel) ranging from $\veps_{1,max}=3.2$ to $\veps_{1,min}=0.2$ (see Eq. (\ref{invsin}, \ref{sin})). Insets: Plots $\omega_{\bb l}^{\rm spp}$, $\omega_{011}^{\rm phot}$ for each case.}
\label{plot1}
\end{figure}

We have also confirmed these findings numerically by assuming both an inverse sinusoidal, cf. Eq. (\ref{invsin}), and a sinusoidal variation, cf. Eq. (\ref{sin}), for region $I$, see upper and lower panels in Fig. \ref{plot1}, respectively. In both examples we have assumed $\veps_{1,max}=3.2$ to $\veps_{1,min}=0.2$ where as we mentioned the inverse profile is meant to model the laser irradiation of a doped semiconductor, while the latter one models the Rabi like oscillations in a pure semiconductor \cite{Uhlmann:2004}. Here we chose the transverse radial section for the dielectric slab and SPPs to be $R_{\rm sp}=R=2.5$ cm, the slab radius (used in $k_\bot=x_{\rm np}/R$). It may be worth mentioning that for $q=(k_\bot c/ \omega_p)\to 0$ (Zenneck waves) the propagation length becomes unbounded, but by enclosing the SM interface within a cavity of transverse section, $k_\bot$ stays bounded.

\par
\begin{figure}[h]
\centering
\scalebox{0.43}{\includegraphics{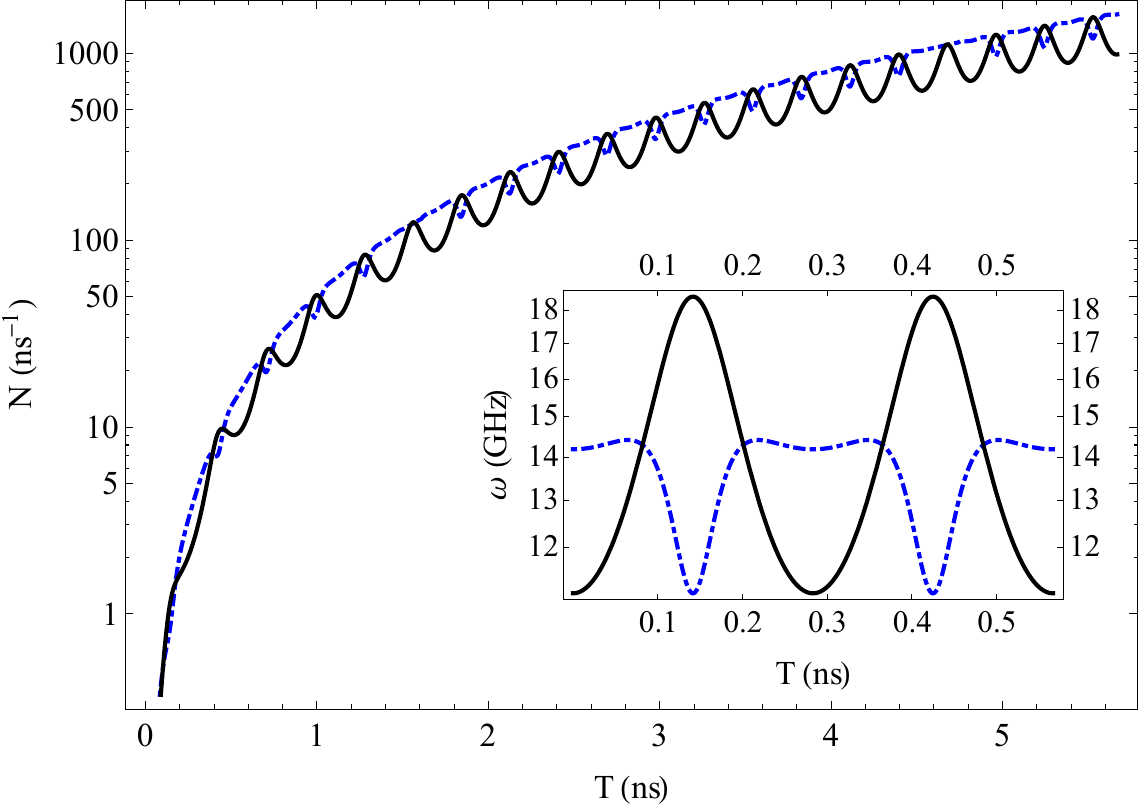}}
\caption{(Color online) Comparison with Fig. \ref{plot1} (lower panel) for TE modes. The particle creation rate for single mode coupling in the instantaneous basis approximation (solid-dark) and the exact solution (dot-dashed-blue) for a sinusoidal variation. cf. Eq. (\ref{sin}), for TE$_{111}$ with the same conditions as in Fig. \ref{plot1}. Inset: Comparison of $\tilde \omega_{111}^{\rm phot}$ and $\omega_{111}^{\rm phot}$.}
\label{plot2}
\end{figure}

 In Fig. \ref{plot2} we also plotted the creation rate numerically for the fundamental TE${}_{111}$ cylindrical mode, for a sinusoidal variation.\footnote{For inverse sinusoidal modulations we find that $\tilde \omega^m_{\bb l}$ becomes imaginary for certain times. Note although the TE frequency is not perturbed at leading order  (for $a\ll L$), the overall factor of $1/\veps(t)$, see Eq. (\ref{wTM}), still leads to shifts in eigenfrequency, cf. Eq. (\ref{tildome}).} We see that the SPP rate is of a similar magnitude for both inverse sinusoidal \& sinusoidal variations and we also see that both TM branch SPP and photons are a magnitude larger when compared to sinusoidal ones. This indicates that using doped-semiconductors would lead to easier detection of pair created photons or vacuum SPPs, although a priori we should consider the effects of dissipation and solve the photon eigenvalues, Eq. (\ref{trans}), for general values of $\eta=a/L$ (see Sec. \ref{conc}).

\section{Detection Scheme} 
\label{detect}

\par In our proposed detection scheme, see Fig. \ref{exp}, we have chosen region $II$ to be that of a metal such as silver and hence satisfies $\veps_2<0$ for frequencies blow the plasma frequency, $\omega_p$, cf. Eq. (\ref{plasma}). To realistically modulate the permittivity, $\veps_1(t)$, we have discussed possible inverse and sinusoidal variations arising from the excitation of localized electrons below the conductions band \cite{Uhlmann:2004} and arising from intra-band transitions in a doped semiconductor, \cite{Haug_Koch}, respectively. As we mentioned, for laser pulses with an energy ($h\lambda/c$) above the band gap, a time varying bulk conductivity, $\rho(t)$, would be generated leading to a modulated permittivity with $\veps_{\rm min}<0$ (shifts $\chi < 0$) and will be left for future investigation. 
Hence, in this article we only consider SM interfaces such that one interface region $I$ has $\veps_1(t)>0$, and the other region $II$ always has $\veps_2<0$. 



\par One possible way to detect vacuum excited SPPs would be to use near-field microscopy with a photon scanning tunneling microscope, e.g., see \cite{Maier}, where the microscope is placed on the opposite vacuum or air side of the SM interface: region $II$, see Fig.~\ref{exp}. 
 Usually, to {\it generate} SPPs a monochromatic light source is sent into a prism placed above the interface with total internal reflection at angle $\theta_i$. The SPPs are  detected by finding a decrease in emitted power at $\theta_i$. However, the time reversed case is equivalent to the creation of vacuum excited SPPs and therefore would lead to a {\it telltale} signature: {\it SPPs would be created via the observation of an increase in emitted power at $\theta_i$ during the time modulation of $\veps_1$}. We should; however, require that the pulsed laser itself does not generate SPPs, as can be arranged by uniformly irradiating the dielectric slab at 90 degrees incidence, see Fig.~\ref{exp}.
\par The experimental details we mentioned so far are simple extensions of current DCE experiments \cite{Agnesi:2009jpc}. However, it may well also be possible to use experiments that have already detected DCE analog radiation in metamaterials \cite{Paraoanu_PNAS110}. The analogy of SPPs in this work with metamaterials comes from considering flux qubits coupled to coplanar waveguides \cite{Paraoanu_PRA82}. Such experimental conditions have already been demonstrated \cite{Astafiev2010} by coupling artificial atoms to carbon nanotubes and it seems within reach of current technology to also adapt these experiments to time-dependent variations of flux qubits (already done in \cite{Wilson:2012, Paraoanu_PNAS110} for photon analogs). 

\par We can also go further with metamaterials, where say $\mu_2 < 0$ where in the late time limit the SPP creation rate is:  
\beq
n_{\bot}^{\rm sp}=\sinh^2\left( {k_\bot^2 c^2}{\kappa \over 4\mu_2 } t \right)
\eeq
for 
\beq
{\mu_2 \over \mu_1(t)} \sim\chi + \kappa \cos(2\omega_{0l} t).
\eeq
Given that $\sinh (-x) = -\sinh x$ there is no problem in having $\mu_2 <0$ because $n_{\bot}^{\rm sp}$ depends quadratically on $\sinh$.

\par This rate is also comparable to the photon-photon rate if $\mu_1$ were varied in time above GHz frequencies. It would be interesting to try and design experiments in centimeter/micrometer sized cavities using split ring resonators and wire rods that are then modulated in time. This leads to easier detection by precisely controlling the SPP wavelength, $\lambda_{\rm sp}$.

\section{Conclusion \& Outlook}
\label{conc}

\par We have discussed how SPPs can be excited out of the vacuum for the case where a dielectric crystal changes from a minimum to maximum value at a semiconductor semispace/metal (SM) interface during laser irradiation. We separated Maxwell's equations in a generalized Lorenz gauge for time-dependent media, both for the permittivity and permeability. For parametric oscillations of a dielectric slab our analytic and numerical analyses show that vacuum excited SPPs can be of the same order of magnitude as the photon-photon rate.

\par The results for the photon creation rate in a cylindrical cavity (generalizing the rectangular case \cite{UhlmanPRL2004}) were also found. For experimental proposals to detect DCE radiation \cite{Agnesi:2009jpc}, the detection of vacuum excited SPPs has added benefits as compared to photon modes: SPPs do not actually need a bounding cavity as they are planar modes (only photon modes need this for parametric enhancement) and for $\mu_{i}=$ constant, TM branch SPP and photon modes are not detuned from their resonant frequencies like TE photons.

\par As future work we should also include dissipative effects, because even without ${\rm Im}[\veps]\neq 0$, it is possible to have imaginary $\tilde\omega$ the photon branch. The separation of variables method used here \cite{PedrosaPRL2009} that we generalized to a Lorenz-like gauge, naturally allows one to incorporate ${\rm Im}[\veps]$.  In fact, for ${\rm Im}[\veps]$, the propagation length via $(2{\rm Im}[k_\bot])^{-1}$ for SPPs diminishes, not the production rate, so this might lead to another benefit. These and other issues including de-tuning of resonant frequencies arising from dissipation will be addressed elsewhere.

\par It would also be interesting to investigate the vacuum excitation of volume/bulk plasmon polaritons (VPPs), usually created by firing a beam of electrons, e.g., see \cite{Raether}, because longitudinal modes are not excited by light and hence require particle impact at an interface, e.g., see \cite{Maier}. However, vacuum excitations might be achieved dynamically by firing clusters of `neutral' Argon atoms at a sample of material \cite{Inui:2008}, or by placing the sample on a high frequency piezo, e.g., see \cite{Kim:2006zzj}. The issue of the gap between bulk and surface plasmons \cite{Raether} indicates they are more difficult to create.

\par Finally, for time dependent media, it has recently been suggested \cite{Bei:2011} that not only transverse (TE and TM) modes, but also longitudinal modes can be created out of vacuum. These are usually {\it unphysical} in Gupta-Bleuler quantization due to a cancellation among time and longitudinal components. However the authors in \cite{Bei:2011} argue that for a time-dependent permittivity: $\veps(t)$, such a cancellation does not occur and surface charges arise as a real physical effect from longitudinal modes. The issue of quantization in general time dependent media requires further investigation, where it would be interesting to find the relationship, or difference,  between vacuum excited SPPs and possible surface charges from longitudinal modes. 

\par  {\it Note Added---} While this work was under revision, a paper dealing with the spontaneous emission of photon pairs from a metamaterial junction \cite{goshmaiti2013} appeared in the literature. However, we consider instead the {\it stimulated emission} of photon pairs from non-adiabatic changes in the vacuum state. We also came across work with similar ideas to those given here: that surface plasmons can be created out of vacuum excitations at a time modulated interface \cite{Hizhnyakov:2015}.

\section*{Acknowledgements}

\par We thank Y.~Kido (Ritsumeikan University), R.~Johansson (RIKEN), G.~S.~Paraoanu (Aalto University School of Science) and H.~Tagawa (Fujitsu) for useful discussions on surface plasmons, quantum circuits, flux qubits and separation of variables, respectively.
\\


\setcounter{equation}{0}
\appendix
\section{Time dependent backgrounds}
\label{timedepMax}

\par Here we discuss a convenient way to separate Maxwell's equations using Hertz vectors. This was developed by Nisbet \cite{Nisbet:1957} for non-dispersive inhomogeneous time-dependent media. Here we generalize to the case of an  isotropic and time-dependent medium.\footnote{Time dependent in both the permittivity and permeability.} 

\par Maxwell's equations in SI units are:
\bea
&&\bm{\nabla} \cdot \bb{B} =0, \qquad\qquad \bm{\nabla}
\times \bb{E} =-\dot \bb{B}\nn
&&\bm{\nabla} \cdot \bb{D} = \rho, \qquad\qquad  
\bm{\nabla} \times
\bb{H} -\dot\bb{D}=\bb J
\label{max}
\eea
where  
\beq
\bb D = \veps(t,\bb x) \bb E \qquad \qquad \bb B = \mu(t,\bb x) \bb H
\eeq
In the above we have assumed both a zero permanent polarization and magnetization ($\bb P_0=\bb M_0=0$) and later we will also assume zero bulk charges and currents ($\rho=0, \,\bb J =0$), although for now we keep them to see how general Maxwell's equations can remain in order to separate them.

\par We now define the electromagnetic fields in terms of gauge potentials as follows:
\beq
\bb B = \bnab\times \bb A~ \qquad\qquad
\bb E = -\partial_t \bb A - \bnab A_0
\eeq
where upon substitution into Maxwell's equations (\ref{max}) we find that Gauss' and Ampere's laws lead to:
\bea
-\bnab \cdot \Big(\veps {\partial\over \partial t}\bb A \Big)-\bnab \cdot \Big( \veps \bnab A_0 \Big)&=&\bb \rho,
 \nonumber\\
 \\
{\partial\over \partial t}\Big({ \veps {\partial\over \partial t} \bb A}\Big)
 +{\partial\over \partial t}\Big(\veps \bnab A_0\Big)+\bm{\nabla} \times \Big({1\over \mu} \bm{\nabla} \times \bb A\Big) &=&\bb J.\nonumber
 \label{gaussamp}
\eea
At this point separation of these coupled equations requires some assumptions to be made. The separation in the Coulomb gauge was achieved in the seminal paper by Dodonov, Klimov and Nikonov \cite{DodonovPRA47} assuming a factorisable ansatz:  $\veps(\bb r,t)=\veps(t)\veps(\bb x)$ and $\mu(\bb r,t)=\mu(t)\mu(\bb x)$. 

\par The separation is more difficult in the Lorenz gauge; however, Nisbet \cite{Nisbet:1957} was able to separate Maxwell's equations using what we shall call a generalized spatial Lorenz gauge:
 \beq
 \mu \veps\partial_t (A_0) + \bm\nabla\cdot (\veps(\bb x) \bb A) = 0,
\label{spatlor}
 \eeq
 assuming time-independent media, which is not the standard Lorenz gauge. 
 For the case of time-dependent media, a generalized temporal Lorenz gauge can be found:
\beq
\mu(t) \partial_t (\veps(t) A_0) + \bm\nabla\cdot \bb A = 0,
\label{lorenz}
\eeq
 which works as long as we assume an isotropic piecewise homogeneous and time dependent media: $\veps(t),\mu(t)$ and $\nabla \mu=\nabla \veps=0$, cf. \cite{Bei:2011} for the case of $\mu=1$. Note a generalized spatio-temporal Lorenz gauge of the form $\mu(t) \partial_t (\veps(t) A_0) + \bm\nabla\cdot (\veps(\bb x) \bb A)$ does not  lead to a complete separation as can be verified. However, the separation of a non-dispersive, inhomogenous, conducting and time-dependent medium assuming a factorisable geometry (cf. \cite{DodonovPRA47}) appears to be possible and will be presented elsewhere.
 
 \par Plugging in the temporal Lorenz gauge, Eq.~(\ref{lorenz}), into Eq.~(\ref{gaussamp}), assuming $\nabla \veps =\nabla \mu =0$, leads to two uncoupled second order differential equations of the form
 \bea
 {\partial\over \partial t} \Big(\mu{\partial\over \partial t}(\veps A_0)\Big) - \nabla^2 A_0 &=& {\rho\over \veps}
 \nn
 \mu{\partial\over \partial t}\Big(\veps {\partial\over \partial t} \bb A\Big)-\nabla^2  \bb A &=&\bb \mu \bb J~,
 \label{waves}
 \eea
 which generalizes the result found in \cite{Bei:2011} when $\mu=1$. The above result also generalizes work in the Coulomb gauge by Dodonov et al. \cite{DodonovPRA47} and work in \cite{UhlmanPRL2004} which considered either time-dependent $\veps$, or $\mu$ using dual potentials and hence does not allow for the inclusion of charge and current densities which break the duality \cite{Jackson} (also see \cite{PedrosaPRL2009} for time-dependent media in the Coulomb gauge). 

\par  Note the two equations (\ref{waves}) are not symmetric in an interchange of $\veps \leftrightarrow \mu$ which is due to the non-trivial time-dependence of the media. However, a symmetric set of equations (with respect to $\veps \leftrightarrow \mu$) can be obtained from the Hertz method as we show in the next section.

\section{Hertz Vectors}
\label{Hertz}

\par We now define two Hertz vectors $\bb\Pi_e$ and $\bb\Pi_m$ as (with $\mu_0=1$)
\beq 
A_0=-{1\over \veps} \bm\nabla\cdot \bb \Pi_e ,\quad
{\bb A}=\mu{\partial \bb \Pi_e \over \partial t} + \bnab\times {\bb \Pi_m}
\eeq
which automatically satisfies the temporal Lorenz gauge condition, Eq.~(\ref{lorenz}), see \cite{Nisbet:1957} for the definition of potentials on a spatial Lorenz gauge.  It is then possible to show that the coupled wave equation, Eq.~(\ref{waves}), separates as:
\\
\bea
\veps(t) \partial_t ({\mu(t) \partial_t  {\bb \Pi}_e}) -\nabla^2 { {\bb \Pi}_e} &=& \bb Q_e
\nn
\mu(t) \partial_t ({\veps(t)\partial_t  {\bb \Pi}_m}) -\nabla^2 { {\bb \Pi}_m} &=& \bb Q_m~,
\label{Maxwell2}
\eea
where TE modes correspond to $\bb \Pi_m$ while TM modes are those for $\bb \Pi_e$ (for more details see below Eq. (\ref{Maxwell})). Here we have included the so-caled ``stream potentials" \cite{Nisbet:1957}:
\bea
\bb\nabla\cdot\bb Q_e=-\rho \nn
\dot{\bb Q}_e+ {1\over \mu} \bb\nabla \times\bb Q_m=\bb J
\eea
set to zero in the main text as we assume that $\rho, \bb J =0$, cf. Eq.~(\ref{Maxwell}). It may also be worth mentioning that these equations are slightly different to the case discussed in \cite{Nesterenko:2012fk} that applies to a dispersive medium: $\veps(\omega),\mu(\omega)$, on a time-independent ($e^{i\omega t}$) background.

\par The electric and magnetic fields can then be written in terms of  Hertz vectors as
\bea
{\bb E}
&=&{1\over \veps} \bnab ( \bnab\cdot \bb \Pi_e ) 
-\partial_t (\mu\,\partial_t \bb \Pi_e ) -\bnab\times {\partial_t\bb \Pi_m}
\nn
&=&{1\over \veps}\bnab\times(\bnab\times {\bb \Pi_e})-\bnab\times {\partial_t\bb \Pi_m}~,
\nn
{\bb B} &=& \mu \bnab\times {\partial\bb \Pi_e\over\partial t}+
\bnab\times(\bnab\times {\bb \Pi_m})~,
\label{fields}
\eea
allowing one to easily isolate TE and TM modes. 

\par For example, TM modes are defined by the parts $\bb E_{TM}, \bb B_{TM}$ coming from $\bb \Pi_e$ 
with $\bb z \cdot \bb B=0$, where a convenient choice of Hertz vectors are: 
$$
{\bb \Pi}_e=\Phi \, \hat \bb z\,,\qquad {\bb \Pi}_m=\Psi \,\hat \bb z 
$$
and $\Phi$ and $\Psi$ represent TM and TE modes respectively. For TM modes we obtain
\bea
\label{TMfields}
{\bb E}_{TM}&=& {1\over \veps}\partial_1\partial_z \Phi 
\unit e_1 +{1\over \veps}\partial_2\partial_z \Phi 
\unit e_2 ~,
\nn\
{\bb B}_{TM} &=& \mu\,\partial_2\partial_t \Phi \unit e_1  -\mu\,\partial_1\partial_t \Phi \unit e_2 
\eea
with a similar expression for TE modes (from $\bb \Pi_m$ 
with $\bb z \cdot \bb E=0$):
\bea
\label{TEfields}
{\bb E}_{TE}&=& -\partial_2\partial_t \Psi 
\unit e_1 + \partial_1\partial_t \Psi 
\unit e_2 ~,
\nn\
{\bb B}_{TE} &=& \partial_1\partial_z \Psi \unit e_1  +\partial_2\partial_z \Psi \unit e_2 ~.
\eea
These generalize the time-independent cases found e.g. in \cite{Nesterenko:2012fk}.

\par These equations of course combine for both TE and TM modes to give the total electric and magnetic field strengths:
\begin{widetext}
\bea
{\bb E}&=& \left({1\over \veps}\partial_1\partial_z \Phi- \partial_2\partial_t\Psi\right) \unit e_1 +\left({1\over \veps}\partial_2\partial_z \Phi+ \partial_1\partial_t\Psi\right) \unit e_2 - {1\over \veps} \bnab_{\bot}^2\Phi \,\unit z
\\
{\bb B} &=& 
\left(\mu\,\partial_2\partial_t \Phi+ \partial_1\partial_z\Psi\right) \unit e_1  + 
 \left(-\mu\,\partial_1\partial_t \Phi+ \partial_2\partial_z \Psi\right) \unit e_2 - \bnab_{\bot}^2\Psi \,\unit z\nonumber
\eea
\end{widetext}
and generalizes the time-independent case, e.g., see \cite{Nesterenko:2012fk}, to the time-dependent case.

\section{Separation of variables in time-dependent media}
\label{instbas}

\par To allow for space-time-dependent mode functions we can also use an instantaneous basis \cite{LawPhysRevA.49.433}: 
\beq
\Phi(\bb r,t)=\sum_m Q_{\bb m}(t) \vp_{\bb m}(\bb r;t)~,
\eeq
where now $t$ becomes a parameter: $\vp(\bb r,t)\to \vp(\bb r; t)$.
The orthonormality again is given by  
\beq
\int_0^{L} dz\, \veps(t)\vp_{\bb m}(\bb r;t)\vp_{\bb n}(\bb r;t)=(\vp_{\bb m},\vp_{\bb n})=\delta_{\bb m\bb n}
\eeq
and satisfies the wave equation:
\beq
\bb\nabla^2 \vp_{\bb m}(\bb r;t)+\veps(t)\mu(t) \omega_{\bb m}^2(t)\vp_{\bb m}(\bb r;t)=0.
\eeq
These steps appear to be identical to the standard separation of variables; however the  time dependent wave equation now becomes \cite{LawPhysRevA.49.433}, cf. Eq. (\ref{TEtimeQ}):
\bea
&\ddot Q_{\bb m} +  \omega_{\bb m}^2 (t) Q_{\bb m}=&
\\
- &\sum_{\bb m}^{\infty} \Big[ 2 M_{\bb {m n}} \dot Q_{\bb n}+\dot M_{\bb {m n}} Q_{\bb n} +  \sum_\ell^\infty M_{\bb {n l}} M_{\bb {m l }} Q_{\bb n} \Big]&
\nonumber\\
&=0 \quad \forall \quad  M_{\bb {m n}} \to 0&
\nonumber
\eea
where the intermode coupling matrix is given by
\beq
{\cal M}_{\bb m\bb n}= \int_0^{L} dz\, \veps(t)\vp_{\bb  m}(\bb r;t)\partial_t \vp_{\bb n}(\bb r;t),
\label{couple}
\eeq\
and for a crystal in free space the bounds would be $\pm\infty$.

\par That is the instantaneous basis approximation assumes that the variable $t$ becomes a parameter, such that we can freeze time derivatives of $\dot \veps=\ddot\veps=0$ (or $\mu$). On the other hand, in the usual separation of variables, cf.  Eq. (\ref{TEtimeQ}), the time derivatives remain, but a rescaling of the mode functions allows us to find a Mathieu like solution, see Eq. (\ref{mathieu}). As we shall see, in the instantaneous approach, instead of these terms we obtain an infinite set of coupled mode equations. 

\par Upon substituting the mode expansion for the instantaneous basis into the Lagrangian density, Eq. (1), 
and then integrating over the spacial part using the orthonormality of the mode functions, defining the conjugate momentum as
\beq
\Pi(\bb r, t) = \veps(t)\sum_{\bb m} P_{\bb m}(t) \vp_{\bb m}(\bb r; t)~,
\eeq
we obtain, via a Legendre transform, a Hamiltonian of the form \cite{Schutzhold:1997yh}:
\beq
H_{\rm eff} = \sum_{\bb m} \left( P_{\bb m}^2+ \omega_{\bb m}^2(t)Q_{\bb m}^2\right)+ \sum_{\bb m \bb n} P_{\bb m} Q_{\bb n} {\cal M}_{\bb m\bb n}(t)~.
\label{Ham}
\eeq
Now the conjugate momentum is defined by 
\beq
P_{\bb m} = \dot Q_{\bb m} - {\cal M}_{\bb m \bb n} Q_{\bb n} ~.
\eeq
Only in special cases does the intermode coupling matrix, Eq. (\ref{couple}), become zero, such as for certain cavity geometries or for a uniform dielectric filling the whole cavity ($a=L$) \cite{LawPhysRevA.49.433}. However, in general,  both methods introduce detuning of the parametric enhancement. In the  separation of variables this comes from the shifted dispersion relation, $\tilde{ \omega}_{\bb l}(t)$, while in the instantaneous basis it comes from the intermode coupling term, ${\cal M}_{\bb m\bb n}(t)$. 

\par Specifically, it is important to note that for SPPs considered here, the definition in Eq. (\ref{plas}) implies there are no intermode coupling terms: ${\cal M}_{\bb m\bb n}=0$ as imposed by the orthonormality of the SPPs. Hence the instantaneous basis leads to a single mode equation which is only identical to the separation of variables approach for $\dot\veps,\ddot\veps =0$. This is exemplified by the fact that the frequencies do not depend on time derivatives of $\veps$ in the instantaneous basis method, cf. Eq. (\ref{TEtimeQ}): $\tilde \omega^{\rm sp}\neq \omega^{\rm sp}$, see the insets in Fig. \ref{plot2}. 


\bibliography{DCE_April2015}

\end{document}